\PassOptionsToPackage{unicode}{hyperref}
\PassOptionsToPackage{hyphens}{url}
\PassOptionsToPackage{dvipsnames,svgnames,x11names}{xcolor}

\documentclass[]{article}

\usepackage[utf8]{inputenc}
\usepackage[T1]{fontenc}
\usepackage{textcomp} 
\usepackage{lmodern}
\usepackage{amsmath,amssymb}
\usepackage{microtype} 
\UseMicrotypeSet[protrusion]{basicmath}
\usepackage{parskip}
\usepackage{xcolor}
\usepackage{graphicx}

\makeatletter
\def\maxwidth{\ifdim\Gin@nat@width>\linewidth\linewidth\else\Gin@nat@width\fi}
\def\maxheight{\ifdim\Gin@nat@height>\textheight\textheight\else\Gin@nat@height\fi}
\makeatother
\setkeys{Gin}{width=\maxwidth,height=\maxheight,keepaspectratio}
\def\fps@figure{htbp}
\NewDocumentCommand\citeproctext{}{}
\NewDocumentCommand\citeproc{mm}{%
  \begingroup\def\citeproctext{#2}\cite{#1}\endgroup}
\makeatletter
\let\@cite@ofmt\@firstofone
\def\@biblabel#1{}
\def\@cite#1#2{{#1\if@tempswa , #2\fi}}
\makeatother

\newlength{\cslhangindent}
\newlength{\csllabelwidth}
\newenvironment{CSLReferences}[2]
 {\begin{list}{}{%
  \setlength{\itemindent}{0pt}
  \setlength{\leftmargin}{0pt}
  \setlength{\parsep}{0pt}
  \ifodd #1
   \setlength{\leftmargin}{\cslhangindent}
   \setlength{\itemindent}{-1\cslhangindent}
  \fi
  \setlength{\itemsep}{#2\baselineskip}}}
 {\end{list}}

\usepackage{calc}

\usepackage[english]{babel}

\IfFileExists{bookmark.sty}{\usepackage{bookmark}}{\usepackage{hyperref}}
\IfFileExists{xurl.sty}{\usepackage{xurl}}{}
\hypersetup{
  pdftitle={peaks: a Python package for analysis of angle-resolved photoemission and related spectroscopies},
  pdfauthor={Phil D. C. King, Brendan Edwards, Shu Mo, Tommaso Antonelli, Edgar Abarca Morales, Lewis Hart, Liam Trzaska},
  pdflang={en-US},
  colorlinks=true,
  linkcolor={Maroon},
  filecolor={Maroon},
  citecolor={Blue},
  urlcolor={Blue},
  pdfcreator={LaTeX via pandoc}}

\usepackage[affil-it]{authblk}
\usepackage{orcidlink}
\title{\texttt{peaks}: a Python package for analysis of angle-resolved photoemission and related spectroscopies}

\author[1]{Phil D. C. King\,\orcidlink{0000-0002-6523-9034}}
\author[1]{Brendan Edwards\,\orcidlink{0000-0002-7219-4241}}
\author[1]{Shu Mo\,\orcidlink{0009-0009-1750-0363}}
\author[1]{Tommaso Antonelli\,\orcidlink{0000-0002-4990-7813}}
\author[1]{Edgar Abarca Morales\,\orcidlink{0000-0002-7714-8228}}
\author[1]{Lewis Hart\,\orcidlink{0000-0002-8598-6272}}
\author[1]{Liam Trzaska\,\orcidlink{0009-0002-1556-3378}}

\affil[1]{SUPA, School of Physics and Astronomy, University of St Andrews, St Andrews, KY16 9SS, United Kingdom}

\date{2 October 2026}

\begin{document}
\maketitle

\section{Summary}\label{summary}

The electronic structure of materials dictates their electrical, optical, and thermodynamic properties. Angle-resolved photoemission spectroscopy (ARPES) provides a direct, momentum-resolved, experimental probe of such electronic structures, and so is widely employed in the study of functional, quantum, and 2D materials
(\citeproc{ref-Damascelli2003}{Damascelli et al., 2003};
\citeproc{ref-king_angle_2021}{King et al., 2021};
\citeproc{ref-sobota_angle-resolved_2021}{Sobota et al., 2021}).
\texttt{peaks} (\textbf{P}ython \textbf{E}lectron spectroscopy
\textbf{A}nalysis by \textbf{K}ing group @ \textbf{S}t Andrews) provides a Python package for advanced data analysis of ARPES and related spectroscopic data. It facilitates the fast visualisation and analysis of multi-dimensional datasets, allows for the complex data hierarchy typical to ARPES experiments, and supports lazy data loading and parallel processing, reflecting the ever-increasing data volumes used in ARPES. It is designed to be run in an interactive notebook environment, with extensive inline and pop-out GUI support for data visualisation.

\section{Statement of need}\label{statement-of-need}

Over recent years, significant technological improvements have developed ARPES into a truly multidimensional spectroscopy. Besides the traditional resolution of energy and up to three momentum directions, temperature, spin, spatial, and temporal-dependent ARPES measurements are becoming increasingly common (\citeproc{ref-king_angle_2021}{King et
al., 2021}; \citeproc{ref-sobota_angle-resolved_2021}{Sobota et al.,
2021}), typically requiring efficient handling and advanced analysis of 3-, 4-, and higher-dimensional datasets. Extensive use of international light sources for performing ARPES measurements, during intensive experiment campaigns running over several days, further motivates a collaborative approach to performing data analysis. There is also an ever-increasing push to incorporate machine learning (ML) methods into the analysis pipeline
(\citeproc{ref-agustsson_autoencoder_2025}{Ágústsson et al., 2025};
\citeproc{ref-iwasawa_unsupervised_2022}{Iwasawa et al., 2022};
\citeproc{ref-kim_deep_2021}{Kim et al., 2021};
\citeproc{ref-melton_k-means-driven_2020}{Melton et al., 2020}), while greater transparency and reproducibility in ARPES data analysis can be ensured by the development and utilisation of open-source packages, with clear and transparent metadata handling
(\citeproc{ref-scheffler_fair_2022}{Scheffler et al., 2022}). The above requirements all motivate the use of Python as a modern approach to ARPES data analysis.

\section{State of the field}\label{state-of-the-field}

Given the above requirements, a number of packages have recently been developed for Python-based ARPES analysis. \texttt{PyARPES}
(\citeproc{ref-stansbury_pyarpes_2020}{Stansbury \& Lanzara, 2020})
represents a pioneering development in this direction. It appears to no longer be actively maintained by the original author, although a maintained fork does exist (\citeproc{ref-pyarpes_fork}{Arafune, 2026}).
Despite many excellent features, it makes several fundamental convention choices (regarding angular and energy scales and units, alignments, and sign conventions) which, in our view, complicate its use when employed with multiple experimental setups as is typical in the ARPES community, while approximations are used in the critical momentum-space conversions. \texttt{pesto} (\citeproc{ref-pesto}{Polley, 2025}) is an excellent easy-to-use alternative, but is heavily oriented towards use with data collected from the Bloch beamline of the Max-IV synchrotron. At the time of writing, we have also recently discovered \texttt{ERLabPy}
(\citeproc{ref-ErLabPy}{Higashihira Han, 2026}) which provides similar functionality to \texttt{peaks}, although with some differences in the approach to handling the data (e.g.~co-ordinate systems). The need to not only accommodate different data formats, but also to manage distinct angle and sign convention choices for data acquired at multiple facilities, can add significant complexity for the user -- in particular for on-the-fly processing. This is something that \texttt{peaks} attempts to simplify for the end user, aiding quick and efficient on-the-fly analysis e.g. for sample alignment during intense experimental runs. Other packages that we are aware of tend to focus on a subset of the functions required, e.g.~for ARPES data analysis (\citeproc{ref-das}{Das, 2026}) or visualisation (\citeproc{ref-Kramer2021}{Kramer \& Chang, 2021}), or are based around GUI-driven workflows (\citeproc{ref-Pudelko2025}{Pudelko et al., 2025}).

\section{\texorpdfstring{\texttt{peaks}}{peaks}}\label{peaks}

\texttt{peaks} attempts to provide a relatively comprehensive suite of tools for ARPES and related spectroscopic data via a modular approach, supporting the experimenter from initial data acquisition, visualisation, and sample alignment through data processing and more advanced analysis. \texttt{peaks} builds heavily on the \texttt{xarray}
package (\citeproc{ref-hoyer2017xarray}{Hoyer \& Hamman, 2017}),
providing a powerful data structure for the N-D labelled data arrays common to ARPES data. This also supports the use of \texttt{dask} arrays
(\citeproc{ref-dask}{Dask Development Team, 2016}) for lazy data loading and processing, e.g.~for datasets that are beyond the available memory, or to facilitate parallel processing. While not a requirement, \texttt{peaks} was intended to be run using interactive notebooks. Data is loaded into
\texttt{xarray.DataArray} objects using location-specific data loaders to support multiple starting data formats and conventions, reflecting the heterogeneity in existing ARPES setups. Extensive metadata is included in the \texttt{DataArray} attributes, making use of \texttt{pydantic}
(\citeproc{ref-Colvin_Pydantic_2025}{Colvin et al., 2026}) models to
ensure a consistent metadata framework, while \texttt{pint}
(\citeproc{ref-pint}{Grecco, 2026}) is used for ensuring reliable handling of units in both the ARPES dataset and associated metadata. Data can also be loaded into \texttt{xarray.DataTree} structures,
allowing the user flexibility over grouping data in configurations which reflect the data hierarchy of the underlying experiment, and permitting batch processing or metadata configuration. General data loaders exist for several of the core ARPES spectrometer manufacturers, as well as for the ARPES setups of several central facilities commonly used in the community. A class-based approach for the data loaders provides an efficient route to extending this to new setups in future. For data saved using the standard data formats of one of the common ARPES spectrometer manufacturers, implementing a new loader can be as simple as subclassing the relevant parent class and defining a few sign and unit conventions, while complete loaders can also be developed starting from bespoke data formats. \texttt{peaks} aims to maintain a record of processing steps that have been applied to the data, building up a detailed analysis history which can be easily inspected. This -- together with use in interactive notebooks -- facilitates enhanced data provenance and effective collaborative working on ARPES data analysis.

\begin{figure}
\centering
\includegraphics[width=0.95\textwidth,height=\textheight]{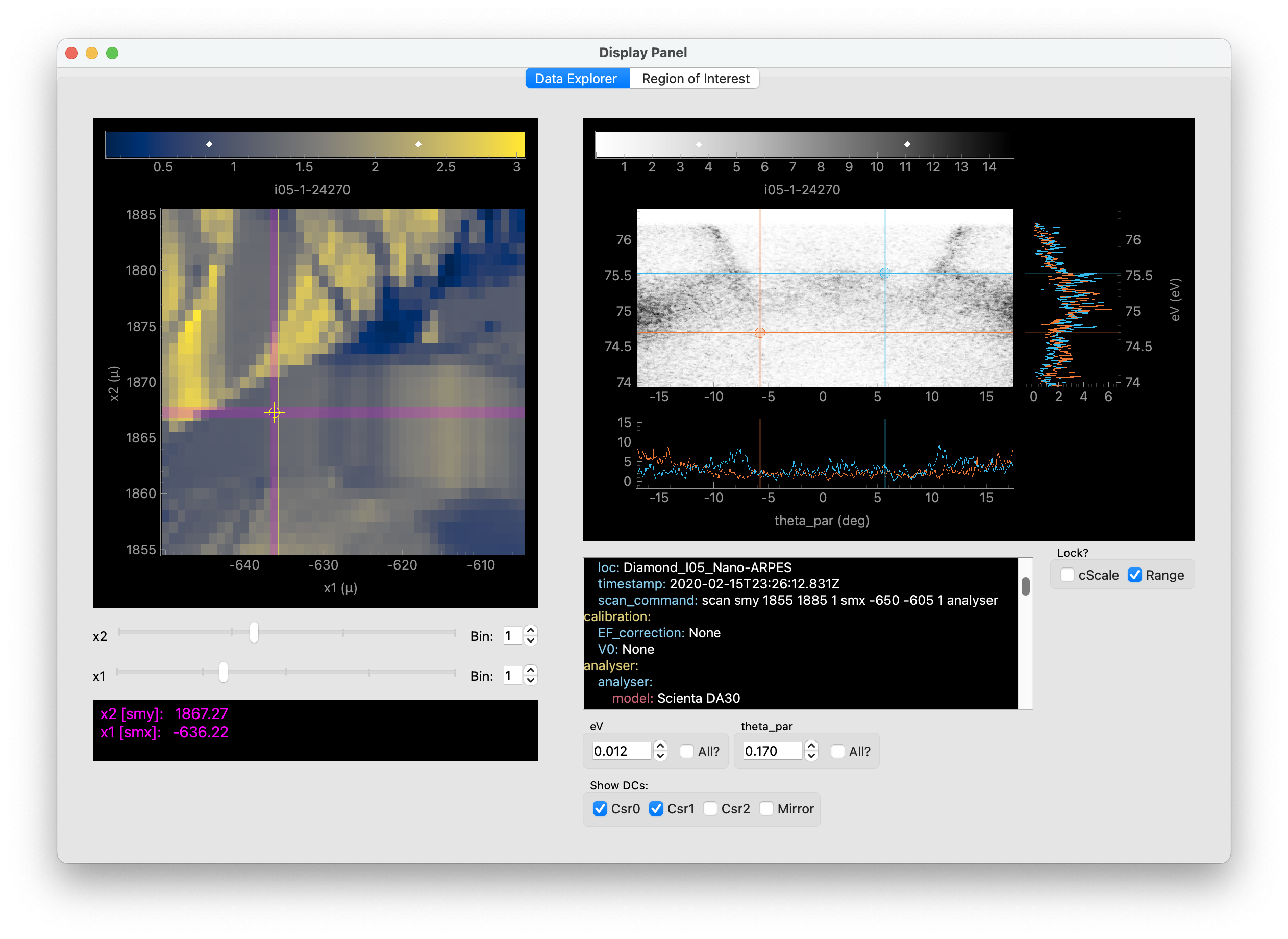}
\caption{Example data visualisation tool for a 4D dataset (spatially-resolved ARPES). The tab shown allows interactive data exploration, while the secondary tab facilitates region of interest analysis.\label{fig:data_viewer}}
\end{figure}

\begin{figure}
\centering
\includegraphics[width=0.75\textwidth,height=\textheight]{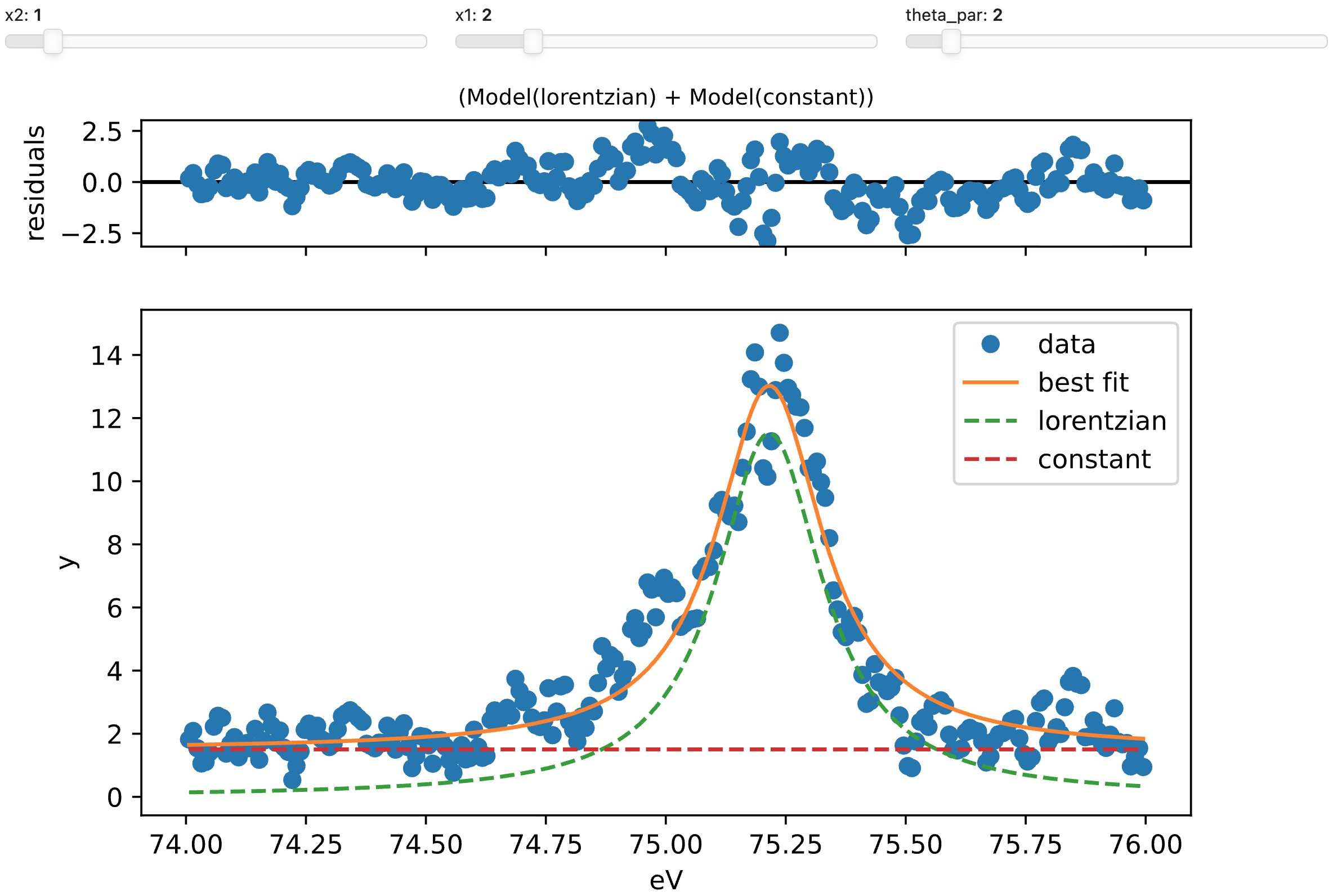}
\caption{Example fitting of lazily-loaded data. The relevant individual EDC is loaded into memory and the fit performed only when selected by the sliders, facilitating fitting even of very large datasets. Similar approaches allow the parallel fitting of individual EDCs or MDCs from across a large, e.g.~spatially-resolved,
dataset.\label{fig:lazy_fitting}}
\end{figure}

The use of \texttt{xarray.DataArray} accessors allows easy chaining of analysis methods together for most functions. Extensive capabilities are included for data visualisation, including static plots and interactive tools for 2-, 3-, and 4-D datasets (see, e.g.~\autoref{fig:data_viewer}). Tools are included for aiding the experimenter in aligning samples for subsequent measurements, with care taken to handle the different conventions used at different experimental facilities in a way that facilitates both standardised data analysis and also the use of \texttt{peaks} for `on-the-fly' analysis during experiments. Additional core functionality includes tools for ARPES-specific data selection (e.g.~momentum (MDC) and energy (EDC) distribution curve extraction), merging, summation, and symmetrisation, data processing (e.g.~momentum conversion and Fermi level corrections, data normalisation), and derivative-type methods to aid data visualisation. Capabilities for data fitting (including parallel processing and fitting of lazily-loaded data - see e.g.~\autoref{fig:lazy_fitting}) are included, building on the extensive
\texttt{lmfit}
(\citeproc{ref-Newville_LMFIT_Non-Linear_Least-Squares_2025}{Newville et
al., 2025}) package. Core capabilities for processing time-resolved ARPES data are included, while specific data selection and helper methods are included for spatially-resolved ARPES. Initial functionality (principal component analysis, clustering, and denoising) is included for related unsupervised machine learning analysis of such spatially-resolved ARPES data, built around the standard Scikit-learn framework (\citeproc{ref-scikit-learn}{Pedregosa et al., 2011}).

In the future, \texttt{peaks} could be augmented with additional ML
approaches tailored to ARPES data analysis, facilitated by the standard
\texttt{xarray}-based data structures used. The incorporation of additional data structures and functionality for processing spin-resolved ARPES data is also planned.

\section{Software design}\label{software-design}

\texttt{peaks} was developed and designed to provide a user-friendly unified platform in response to increasingly complex data acquisition and data processing in the photoemission community. A core aim was to abstract away many of the low-level details related to specific experimental ARPES implementations (e.g. co-ordinate names, angle conventions) to provide a unified experience for data processing, while still allowing for easy referencing to the local conventions as required for performing experiments at multiple facilities. To this end, \texttt{peaks} aims to cover the most common use cases encountered in modern ARPES analysis, from the initial on-the-fly analysis during data acquisition at various facilities worldwide through to the final data presentation for publications.

Being primarily intended for use by experimental physicists, a high-level programming language is required, with Python the natural choice. While the underlying experimental data acquired from the spectrometer is array-like, the varied co-ordinates involved (angles, energies, spatial positions, temperatures, and more) mean that it is crucial to keep this tightly coupled with multiple axis scales and names. The labelled tensor data structure of \texttt{xarray} is an ideal structure for this, and so we utilised this as our fundamental data structure. This has further advantages that it can be used with \texttt{dask} arrays, allowing lazy data loading and parallel processing reflecting the increasing data volumes in ARPES. It also provides a data structure that is similar to legacy software in the field, which has conventionally used the software package \texttt{Igor Pro}. It also has excellent integration with interactive notebooks, which are an ideal platform for use, e.g., during team-based experimental campaigns (as are common in this community) in order to provide a running logbook. Where possible, standard packages were utilised for data analysis (e.g.~peak fitting is based around \texttt{lmfit}), and for enforcing robust data and metadata structures (e.g.~\texttt{pydantic} models).

\texttt{peaks} was originally developed and tested within the authors' research group. At the time, the only available extensive Python ARPES package that was - to our knowledge - openly available was \texttt{PyARPES}, which made a number of fundamentally different choices in regard to the treatment of the experimental data and handling of conventions. Given this, and the design goals of \texttt{peaks} discussed above, a choice was made to develop a new package. Since then, some additional packages have been released, but these also adopt several different conventions to \texttt{peaks} (see \nameref{state-of-the-field} section for more details).

\section{Research impact statement}\label{research-impact-statement}

The development of \texttt{peaks} has so far underpinned data analysis in more than 10 academic publications from the developers’ group (e.g.~\citeproc{ref-armitage_electronic_2025}{Armitage et al., 2025}; \citeproc{ref-buchberger_persistence_2025}{Buchberger et al., 2025}; \citeproc{ref-edwards_giant_2023}{Edwards et al., 2023}; \citeproc{ref-mo_resonant_2026}{Mo et al., 2026}; \citeproc{ref-saika_flat_2025}{Saika et al., 2026}; \citeproc{ref-siemann_dichotomy_2025}{Siemann et al., 2025}), during which time it has undergone continuous and active development and maintenance. Since its public release in 2025, we have been contacted by several groups who are starting to use \texttt{peaks} as their primary analysis driver for ARPES, reflecting increased community adoption (e.g.~\citeproc{ref-peaks_pr3}{AndersSMortensen, 2026}), and so we expect its impact to substantially broaden over the coming years.

To support broader adoption, \texttt{peaks} provides detailed documentation covering installation instructions and a thorough user guide which is based on rendered Jupyter notebook outputs. These are written in a didactic style for users who may be familiar with ARPES data analysis but not Python, while a single-line installation from PyPI, including all required dependencies, makes installation simple for new users. The Jupyter notebooks which make up the documentation user guide are available in the \texttt{peaks} GitHub repository, and can be run locally, with automated download of sample data from an open-source Zenodo repository to facilitate training. These notebooks also serve as a test set for \texttt{peaks}, which are validated in CI for all new contributions made to \texttt{peaks} before they are merged into the main repository branch. The package uses semantic versioning, with tagged commits for each version and corresponding releases on PyPI. Software API and contribution guidelines are included in the documentation, and a standard code style is enforced, facilitating broader contributions - which we strongly welcome.

The \texttt{peaks} package is available at \url{https://github.com/phrgab/peaks}.

\section{AI usage disclosure}\label{AI-usage-disclosure}

Scientific logic and overall software design were performed by the authors. GitHub Copilot and ChatGPT were used for pair programming for some aspects of the code development and refactoring. All AI-generated content was reviewed, tested and validated by the authors. AI was not used for the writing of this paper.

\section{Acknowledgements}\label{acknowledgements}

We acknowledge valuable discussions, suggestions, and bug reports/fixes from Marieke Visscher, Gesa Siemann, Naina Kushwaha, Bruno Saika, Phil Murgatroyd, Anđela Živanović, and Igor Marković. We are grateful to past and present members of the King group at the University of St Andrews and collaborators for measurements of the experimental ARPES data that were utilised in the development of the code. We thank the UK Engineering and Physical Sciences Research Council (Grant Nos. EP/X015556/1, EP/T02108X/1, and EP/R025169/1), the Leverhulme Trust (Grant Nos. RL-2016-006 and RPG-2023-256), and the European Research Council (through the QUESTDO project, 714193) for financial support.

\section*{References}\label{references}
\addcontentsline{toc}{section}{References}

\phantomsection\label{refs}


\begin{CSLReferences}{1}{0}
\bibitem[\citeproctext]{ref-agustsson_autoencoder_2025}
Ágústsson, S. Ý., Haque, M. A., Truong, T. T., Bianchi, M., Klyuchnikov,
N., Mottin, D., Karras, P., \& Hofmann, P. (2025). An autoencoder for
compressing angle-resolved photoemission spectroscopy data. \emph{Mach.
Learn.: Sci. Technol.}, \emph{6}, 015019.
\url{https://doi.org/10.1088/2632-2153/ada8f2}

\bibitem[\citeproctext]{ref-peaks_pr3}
AndersSMortensen. (2026). \emph{Pull Request \#3: Loader for SGM4}.
\url{https://github.com/phrgab/peaks/pull/3}

\bibitem[\citeproctext]{ref-pyarpes_fork}
Arafune, R. (2026). PyARPES corrected (V5). In \emph{GitHub repository}.
GitHub. \url{https://github.com/arafune/arpes}

\bibitem[\citeproctext]{ref-armitage_electronic_2025}
Armitage, O., Kushwaha, N., Rajan, A., Rhodes, L. C., Buchberger, S.,
Saika, B. K., Mo, S., Watson, M. D., King, P. D. C., \& Wahl, P. (2025).
Electronic structure of monolayer CrTe$_2$: An antiferromagnetic
two-dimensional van der Waals material. \emph{Phys. Rev. B}, \emph{112},
245416. \url{https://doi.org/10.1103/h4h8-j473}

\bibitem[\citeproctext]{ref-buchberger_persistence_2025}
Buchberger, S., in 't Veld, Y., Rajan, A., Murgatroyd, P. A. E., Edwards,
B., Saika, B. K., Kushwaha, N., Visscher, M. H., Berges, J., Carbone, D.,
Osiecki, J., Polley, C., Wehling, T., \& King, P. D. C. (2025). Persistence
of Charge Ordering Instability to Coulomb Engineering in the Excitonic
Insulator Candidate TiSe$_2$. \emph{Phys. Rev. X}, \emph{15}, 041028.
\url{https://doi.org/10.1103/9trc-9865}

\bibitem[\citeproctext]{ref-Colvin_Pydantic_2025}
Colvin, S., Jolibois, E., Ramezani, H., Garcia Badaracco, A., Dorsey,
T., Montague, D., Matveenko, S., Trylesinski, M., Runkle, S., Hewitt,
D., Hall, A., \& Plot, V. (2026). \emph{{Pydantic Validation}} (Version v2.13.4).
\url{https://github.com/pydantic/pydantic}

\bibitem[\citeproctext]{ref-Damascelli2003}
Damascelli, A., Hussain, Z., \& Shen, Z.-X. (2003). Angle-resolved
photoemission studies of the cuprate superconductors. \emph{Rev. Mod.
Phys.}, \emph{75}, 473--541.
\url{https://doi.org/10.1103/RevModPhys.75.473}

\bibitem[\citeproctext]{ref-das}
Das, P. (2026). ARPES python tools. In \emph{GitHub repository}. GitHub.
\url{https://github.com/pranabdas/arpespythontools}

\bibitem[\citeproctext]{ref-dask}
Dask Development Team. (2016). \emph{Dask: Library for dynamic task
scheduling}. \url{http://dask.pydata.org}

\bibitem[\citeproctext]{ref-edwards_giant_2023}
Edwards, B., Dowinton, O., Hall, A. E., Murgatroyd, P. A. E., Buchberger,
S., Antonelli, T., Siemann, G.-R., Rajan, A., Morales, E. A., Zivanovic,
A., Bigi, C., Belosludov, R. V., Polley, C. M., Carbone, D., Mayoh, D. A.,
Balakrishnan, G., Bahramy, M. S., \& King, P. D. C. (2023). Giant
valley-Zeeman coupling in the surface layer of an intercalated transition
metal dichalcogenide. \emph{Nat. Mater.}, \emph{22}, 459--465.
\url{https://doi.org/10.1038/s41563-022-01459-z}

\bibitem[\citeproctext]{ref-pint}
Grecco, H. E. (2026). Pint. In \emph{GitHub repository}. GitHub.
\url{https://github.com/hgrecco/pint}

\bibitem[\citeproctext]{ref-ErLabPy}
Higashihira Han, K. (2026). \emph{erlab: A Complete Python Workflow for ARPES Experiments}. Zenodo.
\url{https://doi.org/10.5281/zenodo.18392129}

\bibitem[\citeproctext]{ref-hoyer2017xarray}
Hoyer, S., \& Hamman, J. (2017). xarray: {N-D} labeled arrays and
datasets in {Python}. \emph{Journal of Open Research Software},
\emph{5}(10). \url{https://doi.org/10.5334/jors.148}

\bibitem[\citeproctext]{ref-iwasawa_unsupervised_2022}
Iwasawa, H., Ueno, T., Masui, T., \& Tajima, S. (2022). Unsupervised
clustering for identifying spatial inhomogeneity on local electronic
structures. \emph{npj Quantum Mater.}, \emph{7}, 24.
\url{https://doi.org/10.1038/s41535-021-00407-5}

\bibitem[\citeproctext]{ref-kim_deep_2021}
Kim, Y., Oh, D., Huh, S., Song, D., Jeong, S., Kwon, J., Kim, M., Kim,
D., Ryu, H., Jung, J., Kyung, W., Sohn, B., Lee, S., Hyun, J., Lee, Y.,
Kim, Y., \& Kim, C. (2021). Deep learning-based statistical noise
reduction for multidimensional spectral data. \emph{Rev. Sci.
Instrum.}, \emph{92}(7), 073901.
\url{https://doi.org/10.1063/5.0054920}

\bibitem[\citeproctext]{ref-king_angle_2021}
King, P. D. C., Picozzi, S., Egdell, R. G., \& Panaccione, G. (2021).
Angle, {Spin}, and {Depth} {Resolved} {Photoelectron} {Spectroscopy} on
{Quantum} {Materials}. \emph{Chem. Rev.}, \emph{121}, 2816--2856.
\url{https://doi.org/10.1021/acs.chemrev.0c00616}

\bibitem[\citeproctext]{ref-Kramer2021}
Kramer, K., \& Chang, J. (2021). Visualization of multi-dimensional data
-- the data-slicer package. \emph{J. Open Source Softw.},
\emph{6}(60), 2969. \url{https://doi.org/10.21105/joss.02969}

\bibitem[\citeproctext]{ref-melton_k-means-driven_2020}
Melton, C. N., Noack, M. M., Ohta, T., Beechem, T. E., Robinson, J.,
Zhang, X., Bostwick, A., Jozwiak, C., Koch, R. J., Zwart, P. H.,
Hexemer, A., \& Rotenberg, E. (2020). K-means-driven {Gaussian}
{Process} data collection for angle-resolved photoemission spectroscopy.
\emph{Mach. Learn.: Sci. Technol.}, \emph{1}, 045015.
\url{https://doi.org/10.1088/2632-2153/abab61}

\bibitem[\citeproctext]{ref-mo_resonant_2026}
Mo, S., Kovalenka, K., Buchberger, S., Saika, B. K., Azhar, A., Rajan, A.,
Zivanovic, A., Yao, Y.-C., Belosludov, R. V., Watson, M. D., Bahramy, M.
S., \& King, P. D. C. (2026). Resonant Interlayer Coupling in
NbSe$_2$-Graphite Epitaxial Moiré Superlattices. \emph{Adv. Mater.},
\emph{38}(9), e11262. \url{https://doi.org/10.1002/adma.202511262}

\bibitem[\citeproctext]{ref-Newville_LMFIT_Non-Linear_Least-Squares_2025}
Newville, M., Otten, R., Nelson, A., Stensitzki, T., Ingargiola, A.,
Allan, D., Fox, A., Carter, F., \& Rawlik, M. (2025). \emph{{LMFIT:
Non-Linear Least-Squares Minimization and Curve-Fitting for Python}}
(Version 1.3.3). \url{https://doi.org/10.5281/zenodo.15014437}

\bibitem[\citeproctext]{ref-scikit-learn}
Pedregosa, F., Varoquaux, G., Gramfort, A., Michel, V., Thirion, B.,
Grisel, O., Blondel, M., Prettenhofer, P., Weiss, R., Dubourg, V.,
Vanderplas, J., Passos, A., Cournapeau, D., Brucher, M., Perrot, M., \&
Duchesnay, E. (2011). Scikit-learn: Machine learning in {P}ython.
\emph{J. Mach. Learn. Res.}, \emph{12}, 2825--2830 \url{https://jmlr.csail.mit.edu/papers/v12/pedregosa11a.html}

\bibitem[\citeproctext]{ref-pesto}
Polley, C. (2025). Pesto: Photoemission spectroscopy tools. In
\emph{GitLab repository}. GitLab.
\url{https://gitlab.com/flashingLEDs/pesto}

\bibitem[\citeproctext]{ref-Pudelko2025}
Pudelko, W. R., Kramer, K., Küspert, J., Chang, J., \& Plumb, N. C. (2025). PIVA: Photoemission Interface for Visualization and Analysis. \emph{J. Open Source Softw.},
\emph{10}(114), 9129. \url{https://doi.org/10.21105/joss.09129}

\bibitem[\citeproctext]{ref-saika_flat_2025}
Saika, B. K., Buchberger, S., Mo, S., Rajan, A., Halliday, D., Yao, Y.-C.,
Rhodes, L. C., Decitre, H., Balasubramanian, T., Polley, C., Wahl, P., \&
King, P. D. C. (2026). Reconfigurable flat bands from cooperative moiré and
charge order. \emph{arXiv}.
\url{https://doi.org/10.48550/arXiv.2511.05648}

\bibitem[\citeproctext]{ref-scheffler_fair_2022}
Scheffler, M., Aeschlimann, M., Albrecht, M., Bereau, T., Bungartz,
H.-J., Felser, C., Greiner, M., Groß, A., Koch, C. T., Kremer, K.,
Nagel, W. E., Scheidgen, M., Wöll, C., \& Draxl, C. (2022). {FAIR} data
enabling new horizons for materials research. \emph{Nature},
\emph{604}, 635--642.
\url{https://doi.org/10.1038/s41586-022-04501-x}

\bibitem[\citeproctext]{ref-siemann_dichotomy_2025}
Siemann, G.-R., Murgatroyd, P. A. E., Antonelli, T., Morales, E. A., Khim,
S., Rosner, H., Watson, M. D., Mackenzie, A. P., \& King, P. D. C. (2025).
Dichotomy of electron-phonon interactions in the delafossite PdCoO$_2$:
From weak bulk to polaronic surface coupling. \emph{Phys. Rev. B},
\emph{112}, 085113. \url{https://doi.org/10.1103/5w81-7jtk}

\bibitem[\citeproctext]{ref-sobota_angle-resolved_2021}
Sobota, J. A., He, Y., \& Shen, Z.-X. (2021). Angle-resolved
photoemission studies of quantum materials. \emph{Rev. Mod. Phys.},
\emph{93}, 025006. \url{https://doi.org/10.1103/RevModPhys.93.025006}

\bibitem[\citeproctext]{ref-stansbury_pyarpes_2020}
Stansbury, C., \& Lanzara, A. (2020). {PyARPES}: {An} analysis framework
for multimodal angle-resolved photoemission spectroscopies.
\emph{SoftwareX}, \emph{11}, 100472.
\url{https://doi.org/10.1016/j.softx.2020.100472}

\end{CSLReferences}
\end{document}